# Is Weather Chaotic?[1]

**Name: Aleš Raidl**

**Department of Meteorology, Charles University, V Holešovičkách 2, 180 00 Prague 8,**
E-mail: raidl@mbox.troja.mff.cuni.cz

**Abstract:** The correlation dimension and $K_2$-entropy are estimated from meteorological time-series. The results lead us to claim that seasonal variability of weather is under influence of low dimensional dynamics, whereas changes of weather from day to day are governed by high dimensional system(s). Error-doubling time of this system is less than 3 days. We suggest that the outstanding feature of the weather dynamics is deterministic chaos.

## 1. Introduction

Prediction and predictability are matters of fundamental importance to meteorology. Atmosphere is a considerably complex system containing a wide range of processes linked together by complicated non-linear feedbacks. Some people regard weather forecasting as a main task of meteorology (atmospheric physics). One can watch on TV weather "forecasts" one month or even several seasons ahead. Hardly anybody of laymen asks about success of such long-range prediction. Weather forecasters know that there are meteorological regimes that are more predictable than the others. It was already known to John von Neumann in fifties. In conference in Princeton on the Dynamics of Climate (Pfeffer 1960) he divided the motions of the atmosphere into the three groups:

1. Processes which are determined by the initial conditions – we may extrapolate the initial trends over a short period.

2. Processes which are practically independent to the initial conditions – if we want to forecast such processes we concern with particularities of the atmospheric circulation which will always be present.

3. Between the two extreme cases there is another category of processes – if we are sufficiently far from the initial state so that the details of the initial conditions do not express themselves clearly in what has developed. Therefore we can't calculate backwards the initial conditions from the ultimate conditions.

In our paper, we focus on using time-series of values of air temperature and atmospheric pressure to determine whether or not the atmosphere is chaotic and extent to which atmospheric motion is predictable. The data include the daily mean surface temperature and surface pressure of air over period of 30 years (1 January 1961 – 31 December 1990) observed in the eight Czech meteorological stations (Cheb, Praha – Ruzyně, Hradec Králové, Kuchařovice, Praděd, Holešov, Mošnov, Lysá hora). An approach based on technique developed from deterministic chaos theory is discussed.

## 2. State Space Reconstruction

The key tool used for identification of chaos in time-series is embedding scheme (Packard et al. 1980, Takens 1980, Sauer et al. 1991). This procedure enable us to reconstruct the essential features of dynamics in $m$ dimensional state space from scalar time-series. Let us assume that we have recorded a scalar time-series $x(t_i)$, $i=1,...,N$, at constant sampling time intervals $\Delta t = t_{i+1} - t_i$. We build $m$ dimensional vectors in reconstructed state space as follows

---

$$\vec{X}(t_1) = \{x(t_1), x(t_1 + \tau), \ldots, x(t_1 + (m-1)\tau)\},$$
$$\vdots$$
$$\vec{X}(t_i) = \{x(t_i), x(t_i + \tau), \ldots, x(t_i + (m-1)\tau)\}, \tag{1}$$
$$\vdots$$
$$\vec{X}(t_M) = \{x(t_M), x(t_M + \tau), \ldots, x(t_M + (m-1)\tau)\},$$

where $M=N-(m-1)\tau$, $m$ is so-called embedding dimension and $\tau$ is an appropriate time delay (integer multiple of the sampling time $\Delta t$). Embedding theorem tells us that if the dynamical system evolves on a set of dimension $D$ then sufficient condition for choosing embedding dimension $m$ is $m>2D$. In experimental practice, $D$ is unknown. Therefore, procedure of finding the embedding dimension is to increase $m$ step by step and to compute a fractal dimension (or another invariant) for every value of $m$ until the fractal dimension remains almost constant. At this moment we stop adding other components to the time delay vector. The time delay $\tau$ can be chosen almost arbitrarily for an infinite amount of noise-free data. If the data are noisy and limited in number the value of $\tau$ should guarantee an independence of the co-ordinates of the reconstructed state space. In this study, we select the time delay as the lag at which the autocorrelation function $A(k)$ of the time series $x(t_i)$

$$A(k) = \frac{c_k}{c_0}, \quad k=0,1,\ldots,N\text{-}1 \tag{2}$$

where

$$c_k = \sum_{i=1}^{N-i} \frac{[x(t_i) - \bar{x}][x(t_{i+k}) - \bar{x}]}{N}, \quad \bar{x} = \frac{1}{N}\sum_{i=1}^{N} x(t_i), \quad i=1,\ldots,N,$$

attains the value of $1/e$. We obtained $\tau=63$ days and $\tau=4$ days for temperatures and pressure time-series, respectively. The number of data points $N=10957$ for every of the sixteen time-series. For more sophisticated treatment on embedding procedure see, e.g., Sauer et al. (1991), Abarbanel et al. (1993).

## 3. Correlation dimension

We select correlation dimension $\nu$ as a first quantifier describing properties of a possible climate or weather attractor. Correlation dimension provides the number of active degree of freedom of the system. This dimension was introduced by Grassberger and Procaccia (1983a) because it is easily computable from experimental data. Correlation dimension is defined as follows

$$\nu = \lim_{r \to 0} \lim_{M \to \infty} \frac{\ln C^m(r)}{\ln r}, \tag{3}$$

where $C^m(r)$ is the correlation integral

$$C^m(r) = \frac{1}{(M-\tau_0)(M-\tau_0-1)} \sum_{\substack{i,j=1 \\ |i-j|>\tau_0}}^{M} \theta\left(r - \|\vec{X}(t_i) - \vec{X}(t_j)\|\right) \tag{4}$$

and $\theta(u)$ is Heavisede step function

$$\theta(u) = \begin{cases} 0 & \text{if } u < 0 \\ 1 & \text{if } u \geq 0 \end{cases}$$

$\|\ldots\|$ denotes an appropriate norm in the state space. The cut-off parameter $\tau_0$ is introduced into denominator of Eq. (4) in order to eliminate an anomalous shoulder of correlation integral (see also Theiler 1986). In this contribution, we choose $\tau_0=\tau$.

Extraction $\nu$ directly from Eq. (3) is ineffective, therefore numerous estimators have been developed (for excellent review see Theiler 1990). We use the method suggested by Raidl (1996). This approach is based on searching for a "meaningful scaling region" in a plot of *ln C* versus *ln r*. Then $\nu$ is estimated as a slope of the meaningful scaling region.

Figure 1 shows the *ln C – ln r* diagram for the temperature time-series of Cheb station. Two scaling regions, denoted (A) and (B), are visible in this diagram. The slope of the region (A) oscillates between 20 and 30, whereas the slope of the region (B) converges quickly to the value of 1.7 (see Fig. 2). Analogous results were obtained for the other temperature time-series, with the exception of the time-series of the mountain stations (Praděd, Lysá hora) – in these cases the slope of the region (B) converges to the value between 1.9 and 2.0. The regions (B) disappear after subtraction an annual course of temperature from each daily value.

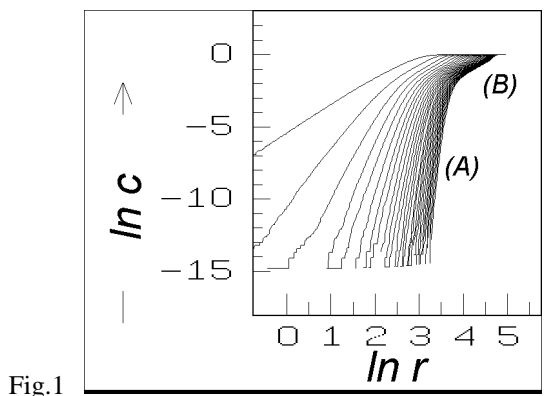
Fig.1

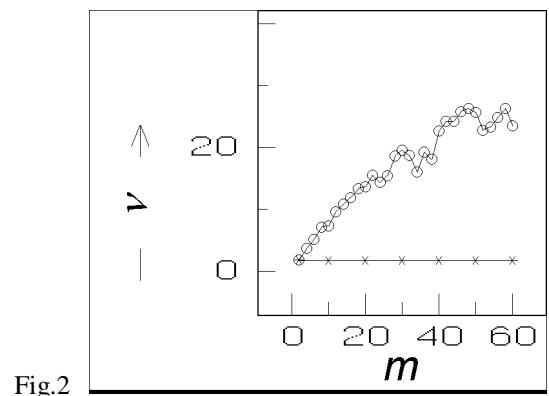
Fig.2

**Figure 1:** *ln C – ln r* diagram for different embedding dimensions *m* (m=2, 4, …, 60) of the temperature time-series of Cheb station.

**Figure 2:** Estimation of correlation dimension $\nu$ as a function of embedding dimension *m* of the temperature time-series of Cheb station.

Figure 3 shows typical correlation integrals of pressure time series. Only one scaling region is apparent in the *ln C – ln r* diagram. Its slope oscillates between 20 and 30 (Fig. 4).

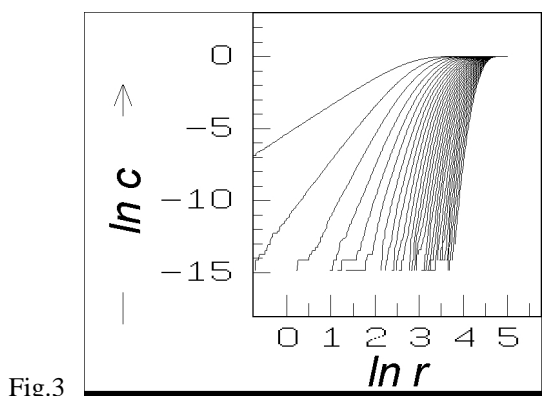
Fig.3

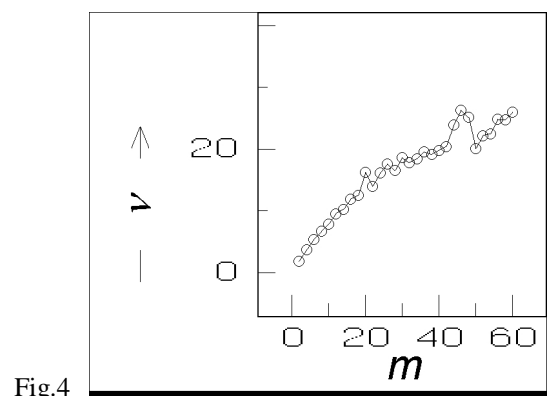
Fig.4

**Figure 3:** *ln C – ln r* diagram for different embedding dimensions *m* (m=2, 4, …, 60) of the pressure time-series of Cheb station.

**Figure 4:** Estimation of correlation dimension $\nu$ as a function of embedding dimension *m* of the pressure time-series of Cheb station.

**4. Entropy**

An non-integer value of $\nu$ does not guarantee that the underlying dynamics is chaotic. Therefore we proceed to estimation of Kolmogorov $K$-entropy. It is well-known (Grassberger and Procaccia 1983b) that:

- $K=0$ in ordered system,

- $K \rightarrow \infty$ in purely random system,

- $0 < K < \infty$ in chaotic (deterministic) system.

We evaluate $K$-entropy through $K_2$-entropy which possesses the following properties: $0 \leq K_2$, $K_2 \leq K$, and has the three above properties of $K$. $K_2$ is easy to compute by means of correlation integrals (Grassberger and Procaccia 1983b):

$$K_2 = \lim_{m \to \infty} \lim_{r \to 0} K_2^m(r), \tag{5}$$

where

$$K_2^m = \frac{1}{k\Delta t} \ln \frac{C^m(r)}{C^{m+k}(r)}, \tag{6}$$

and $k$ is sufficiently small integer number. In order to reduce fluctuations and to improve the statistics, we use averaging of Eq. (6) over five different values of $k$ and approximate dependence of $K_2^m$ on $m$ by means of least-squares fit by the function

$$K_2^m(r) = K_2 + \frac{\nu}{4L\Delta t} \sum_{l=1}^{L} \frac{1}{l} \ln\left(\frac{m+2l}{m}\right), \quad L=5. \tag{7}$$

It stands to reason that $K_2$-entropy has to be evaluated in the same scaling region as $\nu$. Figure 5 depicts $K_2^m$ as a function of $m$ for temperature time-series of Cheb station. The other data provide similar results. The value of the error-doubling time $T_2$ is more illustrative of the predictability of the atmosphere than the value of $K_2$-entropy. The error-doubling time is defined as follows

$$T_2 = \frac{\ln 2}{K_2}. \tag{8}$$

Estimated values of $T_2$ are summarized in Tab. 2. Only the results of extrapolation from scaling regions (A) are mentioned because evaluations from scaling regions (B) give extremely high values of $T_2$ (for more information see Raidl 1995).

|  | Cheb | Holešov | Hradec Králové | Kuchařovic | Mošnov | Ruzyně | Lysá hora | Praděd |
|---|---|---|---|---|---|---|---|---|
| Temperature | 2.9±0.3 | 1.8±0.2 | 1.8±0.1 | 2.1±0.2 | 2.1±0.2 | 2.1±0.3 | 1.9±0.2 | 2.1±0.1 |
| Pressure | 1.3±0.1 | 1.1±0.1 | 1.2±0.1 | 1.2±0.1 | 1.2±0.1 | 1.1±0.1 | 1.1±0.1 | 1.1±0.1 |

**Table 1:** Error-doubling time $T_2$ [day] of the surface temperature and surface pressure time-series.

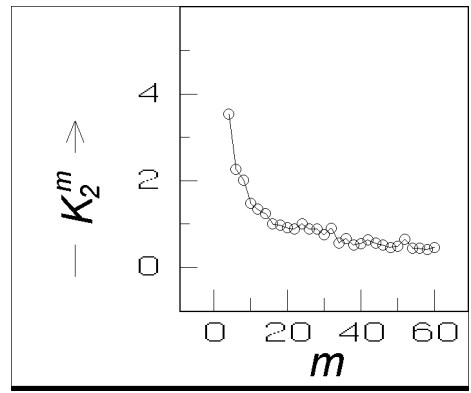

Fig.5

**Figure 5:** Plot of $K_2^m$ as a function of embedding dimension $m$ for the scaling region (A) of the temperature time-series of Cheb station.

## 5. Summary

The ideas of non-linear time series analysis have been used to evaluate correlation dimension and $K_2$-entropy of the atmospheric system. Two scaling regions have been revealed in the correlation integrals of the temperature time series. The first one possessed the gradual slope ($v$ between 1.7 and 2.0) and the second one possessed rapid slope ($v$ between 20 and 30). The first region is linked to the seasonal variability of weather because it disappeared from the $ln\ C - ln\ r$ diagram after removing annual course of temperature from the time-series. Only one scaling region was detected in the correlation integral of the pressure time-series ($v \in \langle 20, 30 \rangle$). We conclude that the seasonal weather variability is governed by a low dimensional dynamical system, whereas the changes of weather from day to day are under the influence of high dimensional dynamics. The estimated values of $K_2$ greater than zero support the suggestion that the dynamics of weather system is chaotic. For that reason the weather is predictable over no more than short time interval ahead. Error-doubling time of the high dimensional system is less than 3 days.

*Acknowledgment.* This work has been supported by the GA ČR under grant 205/98/P140.

## 6. References

Abarbanel H.D.I., R. Brown, J.J. Sidorowich, L. Sh. Tsimring (1993): The analysis of observed chaotic data in physical system, Rev. Mod. Phys., 65, 1331

Grassberger P., I. Procaccia (1983a): Characterization of strange attractors, Phys. Rev. Lett., 50, 346

Grassberger P., I. Procaccia (1983b): Estimation of the Kolmogorov entropy from a chaotic signal, Phys. Rev. A, 28, 2591

Packard N.H., J.P. Crutchifield, J.D. Farmer, R.S. Shaw (1980): Geometry from a time series, Phys. Rev. Lett., 45, 712

Pfeffer R.L. (ed.) (1960): Dynamics of Climate, Pergamon Press, 137 pp.

Raidl A. (1995): Determinism, randomness and predictability of the atmospheric Processes, Ph.D. Thesis, Charles University, Prague, 156 pp. (in Czech)

Raidl A. (1996): Estimating the fractal dimension, $K_2$-entropy and the predictability of the atmosphere, Czech. J. Phys., 46, 293

Sauer T., J.A. Yorke, M. Casdagli (1991): Embodology, J. Stat. Phys., 65, 579

Takens F. (1981): Detecting strange attractors in turbulence, In: Dynamical systems and turbulence (Lecture notes in mathematics), 898, 366, Springer, Berlin

Theiler J. (1986): Spurious dimension from correlation algorithms applied to limited time-series data, Phys. Rev. A, 34, 2427

Theiler J. (1990): Estimating fractal dimension, J. Opt. Soc. Am. A, 1055